\documentclass[]{aa}
\usepackage{graphicx}
\usepackage{txfonts}
\usepackage{natbib}
\usepackage{hyperref} 
\usepackage{dsfont}

\usepackage{amssymb}
\hypersetup{colorlinks=true,linkcolor=blue,citecolor=blue,urlcolor=blue}
\usepackage{orcidlink}
\usepackage{xcolor}

\begin{document}






\title{ Galactic Archaeology with [Mg/Mn] versus [Al/Fe] abundance ratios} 
\subtitle{Uncertainties and caveats}

\author { A. Vasini \orcidlink{0009-0007-0961-0429}\inst{1,2}\thanks{email to: arianna.vasini@inaf.it} ,
E. Spitoni \orcidlink{0000-0001-9715-5727}\inst{2} 
\and 
F. Matteucci  \orcidlink{0000-0001-7067-2302}\inst{1,2,3}}
\institute{  Dipartimento di Fisica, Sezione di Astronomia,
  Universit\`a di Trieste, Via G.~B. Tiepolo 11, I-34143 Trieste,
  Italy \and I.N.A.F. Osservatorio Astronomico di Trieste, via G.B. Tiepolo
 11, 34131, Trieste, Italy  
\and I.N.F.N. Sezione di Trieste, via Valerio 2, 34134 Trieste, Italy
 }

 \date{Received xxxx / Accepted xxxx}

\abstract {
The diagram depicting the abundance ratios  [Mg/Mn] vs. [Al/Fe] has gained significant attention in recent  literature as a valuable tool for exploring 
fundamental aspects of the evolution of the Milky Way and the nearby dwarf galaxies. In particular, this combination of elements is supposed to be highly sensitive to the history of star formation, unveiled by the imprints left on those abundances. Unfortunately, a complete discussion on the uncertainties associated with these nuclei is still missing, making it difficult to know how reliable the associated results are.}{In this paper we analyze, by means of detailed chemical evolution models, the uncertainties related to the nucleosynthesis of Mg, Al, Mn and Fe to show how  different yield prescriptions can substantially affect the trends in the [Mg/Mn] vs. [Al/Fe] plane. In fact, if different nucleosynthesis assumptions produce conflicting results, then the [Mg/Mn] vs. [Al/Fe] 
diagram does not represent a strong diagnostic for the star formation history of a galaxy.}{We discuss the results on the [Mg/Mn] vs. [Al/Fe] diagram, as predicted by several Milky Way  and Large Magellanic Cloud chemical evolution models adopting  different nucleosynthesis prescriptions.}{The results show that the literature nucleosynthesis prescriptions require some corrective factors to reproduce the APOGEE DR17 abundances of Mg, Al and Mn in the Milky Way and that the same factors can also improve the results for the Large Magellanic Cloud. In particular, we show that by modifying the massive stars yields of Mg and Al the behaviour of the [Mg/Mn] vs. [Al/Fe] plot changes substantially  .}{In conclusion, by changing the  chemical yields within their error bars,  one obtains trends which differ significantly, making it very difficult to draw any reliable conclusion on the star formation history of galaxies. The proposed diagram 
is therefore very uncertain from a theoretical point of view and it could represent a good diagnostic for star formation, only if the uncertainties on the nucleosynthesis of the above mentioned elements (Mg, Mn, Al and Fe) could be reduced by future stellar calculations.}


\titlerunning{Galactic Archaeology with [Mg/Mn] versus [Al/Fe]}

\authorrunning{Vasini et al.}

\maketitle

\section{Introduction}
In a recent paper, \citet{fernandes2023} compared the [Mg/Mn] vs. [Al/Fe] relation for the stars of Gaia-Enceladus with those of several dwarf satellite galaxies of the Milky Way, including  LMC and SMC, in order to establish  if the stars in the Galactic halo have been accreted or formed "in situ". The above relation, in fact, is different for different star formation histories, typical of different galactic systems that could have been the building blocks of the halo.

\begin{table*}
\centering
\tiny
\caption{Parameters of the five MW and two LMC models tested (indicated in the first column). The following columns list, in order, the star formation efficiency (time dependent in both cases), the IMF adopted, the timescale of the infall, the wind efficiency (if included in the model), the fraction of Type Ia SNe and the references where to find the nucleosynthesis prescriptions.
}
\begin{tabular}{cccccc}
\hline
 & & & & &\\
 \textbf{Model} & \textbf{Star formation} & \textbf{IMF} & \textbf{Infall rate} & \textbf{Mass loading} & \textbf{Nucleosynthesis}\\
 &  \textbf{efficiency [Gyr$^{-1}$]} & & \textbf{timescale [Gyr]} & \textbf{factor [Gyr$^{-1}$]} & \\
 & & & & $\omega_{i}$ &\\
 & & & & &\\
 \hline
 & & & & &\\
 REF MW & 2.0 (t$<$1 Gyr) & \citet{Kroupa1993} & $\tau_{T}$ = 0.7 & no wind & see Sec. \ref{sec:yields} \\
 & 1.0 (t$\ge$1 Gyr) & & $\tau_{D}$(8 kpc) = 7.0 & & \\
 & & & & &\\
 \hline
 & & & & &\\
 NEW MW & " & " & " & " & see Table \ref{tab:yields_correction} \\
 & & & & &\\
 \hline
 & & & & &\\
 MIX MW & " & " & " & " & see Table \ref{tab:yields_MIX} \\
 & & & & &\\
 \hline
 & & & & &\\
 CL04 MW & " & " & " & " & see Section \ref{sec:MW_andrews} \\
 & & & & &\\
 \hline
 & & & & &\\
 K01 MW & " & \citet{Kroupa01} & " & " & same as REF MW \\
 & & & & &\\
 \hline
 & & & & &\\
 WIND MW & " & \citet{Kroupa1993} & " & 2.5 & same as NEW MW \\
 & & & & &\\
 \hline
 & & & & &\\
 REF LMC & 0.03 (t$<$10.8 Gyr) & \citet{Salpeter55} & $\tau$ = 5.0 & 0.25 & see Sec. \ref{sec:yields}\\
 & 0.25 (10.8 Gyr $<$ t $<$ 11.8 Gyr) & & & &\\
 & 0.1 (11.8 Gyr$<$ t $<$ 13.2 Gyr) & & & &\\
 & 0.4 (13.2 Gyr $<$ t $<$ 13.4 Gyr) & & & &\\
 & 0.35 (13.4 Gyr $<$ t $<$ 13.5 Gyr) & & & &\\
 & 0.8 (t $>$ 13.5 Gyr) & & & &\\
 & & & & &\\
  \hline
 & & & & &\\
 NEW LMC  & " & " & " & " & see Table \ref{tab:yields_correction} \\
 & & & & &\\
 \hline
 & & & & &\\
 MIX LMC & " & " & " & " & see Table \ref{tab:yields_MIX} \\
 & & & & &\\
 \hline
\end{tabular}
\label{tab:model_parameters}
\end{table*}
Galaxy mergers were common in the early universe, and in the last years remnants of merging activity of the Milky Way have been identified in several stellar streams. Moreover, the discovery of Gaia-Enceladus \citep{helmi2018} has suggested that a major merger has occurred roughly 10 Gyr ago \citep{helmi2018,monta2021} in the Galactic halo. The proposed [Mg/Mn] vs. [Al/Fe] relation \citep{Hawkins2015} can in principle discriminate between different star formation rates, and this is because Mn is an element produced mainly in supernovae Type Ia (SNe Ia) with long time delays, and more than Fe which is  also partly produced by massive stars, whereas Mg and Al are mainly produced  by massive stars (supernovae core-collapse, CC-SNe).  The data adopted by \citet{fernandes2023} belong to the DR17 of the APOGEE Survey-2 \citep{Majewski:2017ip,Ahumada2019,apogeedr172022}. In the observational diagram, the stellar systems formed by means of an intense burst of star formation are expected to occupy a particular region with high [Mg/Mn] at high [Al/Fe], while objects evolved with a mild star formation rate, such as dwarf spheroidals, show a lower [Mg/Mn] at low [Al/Fe]. Several observational investigations have indeed revealed a significant number of kinematically identified accreted stellar populations in the Milky Way clumps, in the same region of the above-mentioned chemical plane \citep[i.e.] []{feuillet2021, perottoni2022, horta2023, feltzing2023}.
Therefore, this diagram allows us to discriminate the stars formed in situ from those accreted, better than using the classical [$\alpha$/Fe] versus [Fe/H] ratios, which are very similar in all systems at low metallicity ([Fe/H]), due to the dominance of the enrichment by CC-SNe at early times \citep{spitoni2016}. In fact, these [$\alpha$/Fe] ratios observed in  dwarf spheroidal galaxies (dSphs) and ultra-faint dwarf galaxies (UfDs) can coincide with those of halo stars at extremely low metallicities, while they diverge towards higher metallicities \citep{das2020}. 
The choice of the  [Mg/Mn] ratio is to highlight the $\alpha$-poor population better than [Mg/Fe] does, since Mn is a better tracer of SNeIa. 
It should be noted that in \citet{spitoni2016}, it was suggested to adopt the [Ba/Fe] ratio as a discriminant of accreted stars, instead of the [$\alpha$/Fe] one, since Ba is mainly a s-process element produced by low-mass stars and this ratio can be quite different in different star formation environments, especially at low metallicities. The same technique was also proposed from the observational point of view by \citet{Tolstoy+09} and references therein, where the ratio [Be/Fe] is presented in the case of four different dwarf spheroidals that shows different behaviour due to the SFH of the environment in analysis.

In this article, we show that the theoretical [Mg/Mn] vs. [Al/Fe] relation is strongly model dependent and we highlight how different chemical evolution models can provide different solutions, thus suggesting caution in adopting such a tool, which in any case should be combined with dynamical information, such as eccentricity, energy-angular momentum relation, radial action-angular momentum and action diamond \citep[i.e.,][]{ myeong2019,recioDR32022b,carrillo2023}, before drawing firm conclusions.

In Section \ref{sec:models}, we present chemical models for the Milky Way and for LMC. In Section \ref{sect:data}, the observational samples used in our analysis are described.
In Section \ref{sec:results}, we show our results.  Finally, in Section \ref{conclu_sec}, we critically discuss our results and draw some conclusions. In particular, we discuss the role of the initial mass function (IMF) and the stellar yields on the results. The IMF has been derived only for the Milky Way, so for the satellites of the Galaxy, we can only guess the best one that reproduces the observed features. Concerning the yields, still large uncertainties are present in the yields of Mn and Al.


\begin{figure}
    \centering
    \includegraphics[scale=0.9]{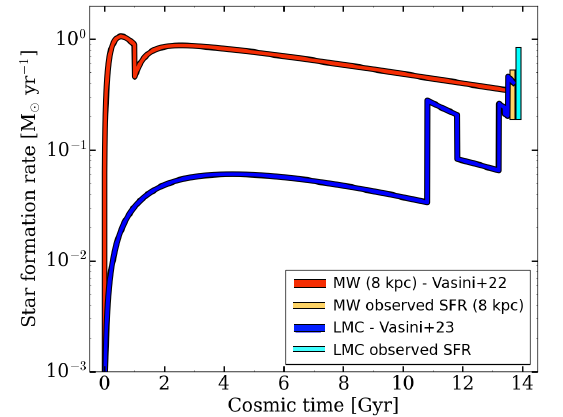}
    \caption{Star formation histories predicted by \citet{Vasini+22,Vasini+23} and used in the paper. For the Milky Way (red line)  the \citet{Kennicutt1998} law with exponent $k=1.5$ has been adopted, and the temporal evolution of the star formation  at the solar ring (8 kpc) is compared  with the present-day observed SFR (vertical yellow line, data taken from \citealt{Prantzos+18}). For the Large Magellanic Cloud (blue line) we assumed a linear \citet{Kennicutt1998} law with a variable efficiency to reproduce the \citet{HZ09} SFR. We compared it with the observation at the present time by \citet{HZ09} itself (vertical cyan line).} 
    \label{fig:MW_LMC_SFR}
\end{figure}

\section{Chemical evolution models for the Milky Way and LMC}

\label{sec:models}
In this paper, we consider as reference models for the Milky Way and for the LMC the ones  of \citet{Vasini+22} and \citet{Vasini+23} respectively, and  we label them as REF MW and REF LMC models. We also introduce NEW MW and NEW LMC models where we considered additional corrective factors on the nucleosynthesis yields, while keeping the other parameters the same. In addition, we combined some reference yields with some corrected ones in the MIX MW and MIX LMC models (see Sec. \ref{subsec:MW_MgMn}). Moreover, we also test three additional MW models with different IMF, nucleosynthesis and wind prescriptions. We present the model parameters in Table \ref{tab:model_parameters}.

In Sec. \ref{sec:model_MW} and Sec. \ref{sec:model_LMC} we show all the prescriptions for the models except the nucleosynthesis assumptions, which are presented later in Sec. \ref{sec:yields}.
\begin{table*}
\centering
\tiny
\caption{Modified  nucleosynthesis yields. Correction factors adopted in the NEW MW and NEW LMC models  for  Fe, Mg,  Al and Mn  considering the same  stellar sources as in \citet{Vasini+22,Vasini+23}. These same prescriptions were used also in the WIND MW model.}
\begin{tabular}{c|c|c|c}
\hline
\multicolumn{4}{c}{}\\
\multicolumn{4}{c}{\textbf{Prescriptions of NEW MW and NEW LMC models}} \\
\multicolumn{4}{c}{}\\
\hline
 & & & \\
 Element & LIMS & Massive stars & SNIa \\
 & \citep{Karakas10} & \citep{Kobayashi+06} & \citep{Iwamoto+99} \\
 & & & \\
 \hline
 Fe & \checkmark & $\times$ 0.7 & \checkmark \\
 \hline
 Mg & \checkmark & Z $\in$ (3-8)$\times$10$^{-3}$ $\longrightarrow$ $\times$1.2 & \checkmark \\
 & & Z > 8$\times$10$^{-3}$ $\longrightarrow$ $\times$ (Z/0.004)$^{0.4}$ & \\
 \hline
 Al &  Z > 8$\times$10$^{-3}$ $\longrightarrow$ $\times$0.8 & $\times$1.5 & \checkmark \\
 \hline
 Mn & \checkmark & \checkmark & $\times$(Z/Z$_{\odot}$)$^{0.65}$ (from \citealt{Cescutti+08})  \\
 \hline
\end{tabular}
\label{tab:yields_correction}
\end{table*}

\begin{table*}
\centering
\tiny
\caption{Yields correction adopted in MIX MW model and MIX LMC model. In these two models we adopted the corrections already introduced in Table \ref{tab:yields_correction} for Mg, Fe and Al. Regarding Mn we adopted the original prescriptions since, as displayed in the right panel of Fig. \ref{fig:MW_abb}, they are the best to reproduce the observed abundance ratio.
}
\begin{tabular}{c|c|c|c}
\hline
\multicolumn{4}{c}{}\\
\multicolumn{4}{c}{\textbf{Prescriptions of MIX MW and MIX LMC models}} \\
\multicolumn{4}{c}{}\\
\hline
 & & & \\
 Element & LIMS & Massive stars & SNIa \\
 & \citep{Karakas10} & \citep{Kobayashi+06} & \citep{Iwamoto+99} \\
 & & & \\
 \hline
 Fe & \checkmark & $\times$ 0.7 & \checkmark \\
 \hline
 Mg & \checkmark & Z $\in$ (3-8)$\times$10$^{-3}$ $\longrightarrow$ $\times$1.2 & \checkmark \\
 & & Z > 8$\times$10$^{-3}$ $\longrightarrow$ $\times$ (Z/0.004)$^{0.4}$ & \\
 \hline
 Al &  Z > 8$\times$10$^{-3}$ $\longrightarrow$ $\times$0.8 & $\times$1.5 & \checkmark \\
 \hline
 Mn & \checkmark & \checkmark & \checkmark  \\
 \hline
\end{tabular}
\label{tab:yields_MIX}
\end{table*}

\begin{figure*}
\centering
\includegraphics[scale=0.7]{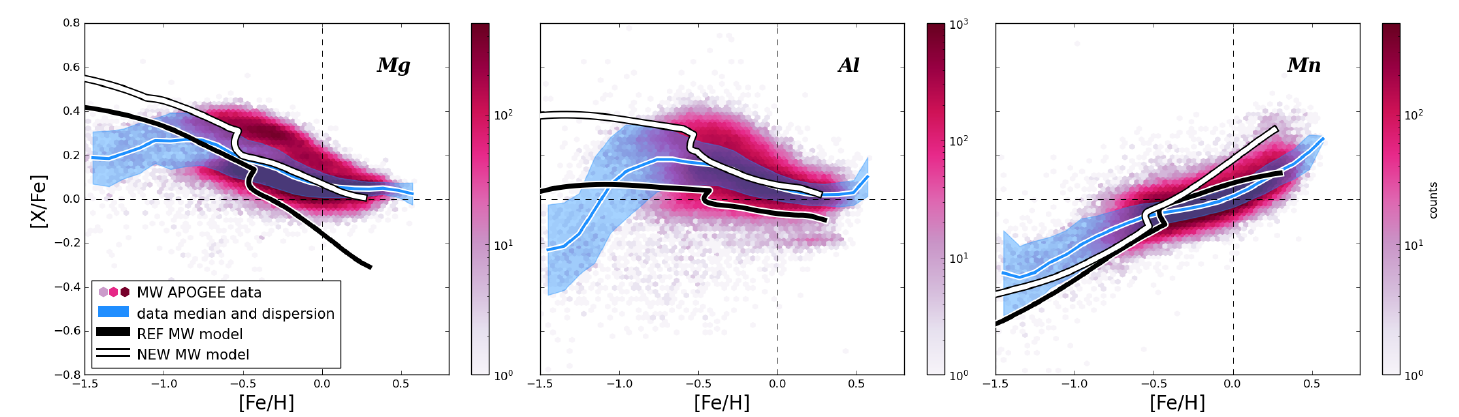}
\caption{Here we show the abundance ratios of the three elements involved. The data are from APOGEE DR17 and are binned and color coded according to the data density in each bin. The blue line represent the running median, computed assuming a bin size of 0.1 dex  and an overlap of 0.025 dex, and the blue shaded area represents the related standard deviation. \textit{Left panel}: [Mg/Fe] in the MW as predicted by REF MW model (black line) and by NEW MW model (white line) compared to the APOGEE data. \textit{Middle panel}: [Al/Fe] MW abundance for the two models (same color coding as Left panel). \textit{Right panel}: [Mn/Fe] abundance for REF MW and NEW MW models (same color coding).}

\label{fig:MW_abb}
\end{figure*}

\subsection{The Milky Way discs}
\label{sec:model_MW}
We adopt the model of \citet{Vasini+22} (where a more extensive description can be found) as a reference for studying the evolution of [Mg/Mn] versus [Al/Fe] abundance ratios in the Milky Way disc components in the solar neighbourhood. This model, originally proposed by \citet{chiappini1997,chiappini2001} and later refined by \citet{Romano&Matteucci03},  \citet{grisoni2018} and \citet{spitoni2019}, is known as the “two-infall model''. 
The main chemical evolution equation on which the model is based, accounts for the chemical enrichment and depletion of the gas:

%


\begin{equation}\begin{split}
\frac{d(X_{\rm i}M_{\rm gas})}{dt}=-X_{\rm i}(t)\psi (t)
+R_{\rm i}(t)+X_{\rm i,A}A(t).
    \label{eq:CEeq}
\end{split}\end{equation}

Here $X_{i}(t)$ and $\psi (t)$ are the abundance by mass of element $i$ and the star formation rate, respectively; $R_{\rm i}(t)$ is the rate at which the element $i$ is restored into the insterstellar medium (ISM); $A(t)$ is the infall rate and $X_{i,A}(t)$ is the abundance of the element $i$ in the infalling gas. 
This model traces the evolution of both the thick and thin discs; the thin disc is  divided into concentric annuli 2 kpc wide, without exchange of matter.  The thick disc is instead considered as a one-zone. It should be noted that in this paper we performed all the calculations at the solar neighborhood (the annular area centred at 8 kpc from the Galactic centre). The formation process involves two distinct gas infall episodes, both following an exponential law. The first accretion event, characterised by a relatively short time-scale ($\tau_T=0.7$ Gyr), leads to the formation of the thick disc. Subsequently, the second accretion event, with a delay from the first one of 1 Gyr, contributes to the formation of the thin disc, which takes place over significantly longer timescales. In particular, it is usually assumed that the thin disk formation timescale depends on the Galactocentric distance, so that an “inside out'' scenario  \citep{matteucci1989} can be reproduced. This dependence follows the law $\tau_{D}(R)=1.033\cdot(R/ \text{kpc})-1.267$ Gyr, as shown by \citet{chiappini2001}.

The star formation rate (SFR) is the \citet{Kennicutt1998} law
\begin{equation}
\psi \propto \nu \sigma_g^k,
\end{equation}
where $\nu$ is the star formation efficiency (fixed at the values of $\nu_T=2$ Gyr$^{-1}$ and $\nu_D=1$ Gyr$^{-1}$ for the thick and thin disc phases, respectively),  $\sigma_g$ is the surface gas density, and $k=1.5$.  
In Fig. \ref{fig:MW_LMC_SFR} we show the star formation history of the Milky Way for the solar neighbourhood (red line) compared to the measured present day as suggested by \citet{Prantzos+18} (in the range 0.2 -- 0.5 M$_{\odot}$ yr$^{-1}$, vertical orange line). In particular, we predict a present day value at the solar neighborhood of $\sim$ 0.3 M$_{\odot}$ yr$^{-1}$.
For the IMF we use that of \citet[][constant in time]{Kroupa1993}. The model accounts for the chemical contribution by SNII, Ib, Ic, Ia as well as novae. The rate for the explosions of these stars are the same used in \citet{Vasini+22}. 

\subsection{The Large Magellanic Cloud}
\label{sec:model_LMC}

To study the chemical evolution of the LMC we adopt the prescriptions of Model 3 presented in \citet{Vasini+23}. The evolution of the abundances in the gas is described by Eq. \ref{eq:CEeq} with the term $-X_{i}(t)W_{i}(t)$ added on the right side of the expression. This accounts for the depletion of gas due to the galactic wind. The wind depends on the SFR $\psi(t)$ through the mass loading factor $\omega_{i}$, whose value is reported in Table \ref{tab:model_parameters}.
In particular the wind rate for the element $i$ is:
\begin{equation}
W_i(t)=\omega_i \psi(t).    
\end{equation}

The chemical evolution model of the LMC is a one-zone model based on a single infall law, a SFR with a linear \citet{Kennicutt1998} law ($k$=1) and a variable efficiency $\nu$ so as to reproduce the recent bursts studied by \citet{HZ09}. This irregular behaviour of the SFR is typical of the structures identified as dwarf irregular galaxies, as the LMC is traditionally classified. This star formation history is shown in Fig. \ref{fig:MW_LMC_SFR} (blue line) compared to the measurements by \citet{HZ09} at the present time (within the interval 0.2 -- 0.8 M$_{\odot}$ yr$^{-1}$, vertical cyan line). We predict a present time SFR of $\sim$ 0.4 M$_{\odot}$ yr$^{-1}$, compatible with the observed range.

\subsection{Nucleosynthesis prescriptions}
\label{sec:yields}

From now on,  we will refer to models REF MW  and REF LMC when the  nucleosynthesis prescription from \citet{Vasini+22} and \citet{Vasini+23} have been considered, which in turn were taken from Model 15 by \citet{Romano+10}. In particular, we adopted \citet{Karakas10} for the nucleosynthesis in AGB stars, \citet{Kobayashi+06} for massive stars, \citet{Iwamoto+99} for SNIa and \citet{JH98} for nova systems. 
\newline

 On the other hand, models NEW MW and NEW LMC are based on the same yields but with additional correction factors tailored to fit the abundance patterns of Mg, Al and Mn in the Milky Way, as explained in Section \ref{subsec:MW_abb}. We also computed a MIX MW model and a MIX LMC model with the combined prescriptions from the other models. The corrections are all listed in Table \ref{tab:yields_correction}, where the first column indicates the element analysed, whereas the second, the third and the fourth are dedicated to Low-Intermediate Mass Stars, Massive stars and SNIa, respectively. The corrections listed are chosen to fit the MW APOGEE DR17 data and are applied to the yields adopted (whose reference is indicated in the first row as a reminder). The only exception is the Mn case, that we corrected as suggested by \citet{Cescutti+08}, who introduced a metal dependency on the yields of this element.
 Then, in the model MW CL04
 we performed a test using nucleosynthesis prescriptions identical to those employed by the chemical evolution models proposed by \citet{andrews2017}. In the same way we performed two more tests with the K01 MW and WIND MW models, assuming the IMF and the wind prescriptions proposed by \citet{andrews2017}, respectively. These prescriptions were chosen since they are the reference ones in the paper by \citet{fernandes2023}.  

\begin{figure*}
    \centering
    \includegraphics[scale=1.15]{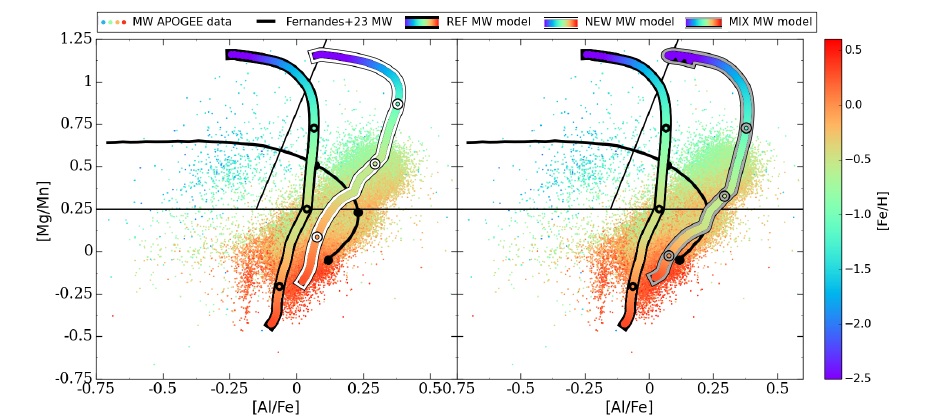}
    \caption{In the two panels we show the data compared to our models and to previous results form the literature. The color coding of the models and the data show the [Fe/H] content as explained by the color bar. In this case, the data are binned so as to show the mean value of [Fe/H] in each bin. The three circles plotted on the models mark the ages 0.3 Gyr, 1 Gyr and 5 Gyr. The horizontal and diagonal black lines separate three different plane regions as denoted by \citet{fernandes2023}: the unevolved populations lie in the upper left region, the MW high-$\alpha$ in situ population lies in the upper right region and the MW low-$\alpha$ in situ population in the lower region. \textit{Left panel}: [Mg/Mn] vs. [Fe/H] for the REF MW model (color coded black-edge line) and for the NEW MW model (color coded white-edge line) compared to the APOGEE DR17 data of \citet{fernandes2023} and their best model obtained adopting the prescriptions of \citet{andrews2017} (black thin line).
     \textit{Right panel}: [Mg/Mn] vs. [Fe/H] for the REF MW model (color coded black-edge line) and for the MIX MW model (color coded grey-edge line) compared to the same data and model by \citet{fernandes2023}.}
    \label{fig:2_MW_MgMn}
\end{figure*}

\section{Observational data}
\label{sect:data}
We compare our model predictions in the [Mg/Mn] versus [Al/Fe] abundance space for the MW and LMC with APOGEE-2 DR17 data  as selected by \citet{fernandes2023}.
The samples discarded  unreliable parameters using the STARFLAG or ASPCAPFLAG={\bf BAD} indicators (see  \citealt{holtzman2015} for definitions). Only  red giant stars have been considered by imposing as conditions: stellar effective temperatures ($T_{\text{eff}}$) ranging from 3750 to 5500 K, and surface gravity log(g)< 3.0 dex.

APOGEE stars of the MW have been selected  using orbital parameters in    the integral of motion considering   circular and  prograde orbits: $L_z > 0$, eccentricity $< 0.3$, $S/N >70$. 
As in \citet{fernandes2023}, we do not impose any condition on the Galactocentric distances.  In fact, model predictions in the solar vicinity (see Section \ref{sec:model_MW}) should well represent the average behaviour of the whole Milky Way disc. They considered bright and faint red giant branch (RGB) stellar populations. In their Table 2, they  summarised the LMC members selection indicating  the sky position, projected distance on the sky, Gaia proper motions, radial velocities, and magnitudes. This selection mimics that proposed by \citet{Nidever+20}, where Table 1 lists the APOGEE MC Fields.

\section{Results}

\label{sec:results}
In this Section, we present model results obtained for the Milky Way and the LMC. First, in Section \ref{subsec:MW_abb} we show the plots for the single abundances ([Mg/Fe], [Al/Fe], and [Mn/Fe]) for the MW and then the trends in the plane [Mg/Mn] vs. [Al/Fe] in Section \ref{subsec:MW_MgMn}. In Section \ref{sec:abb_LMC} and \ref{sec:LMC_MgMn}, we show the abundances and the [Mg/Mn] vs [Al/Fe] trends for the LMC, respectively. Finally, in Section \ref{sec:MW_andrews}, we  test the effects  of adopting the nucleosynthesis prescriptions adopted by \citet{andrews2017} on the chemical evolution of the MW discs.

\begin{figure*}
\centering
\includegraphics[scale=0.7]{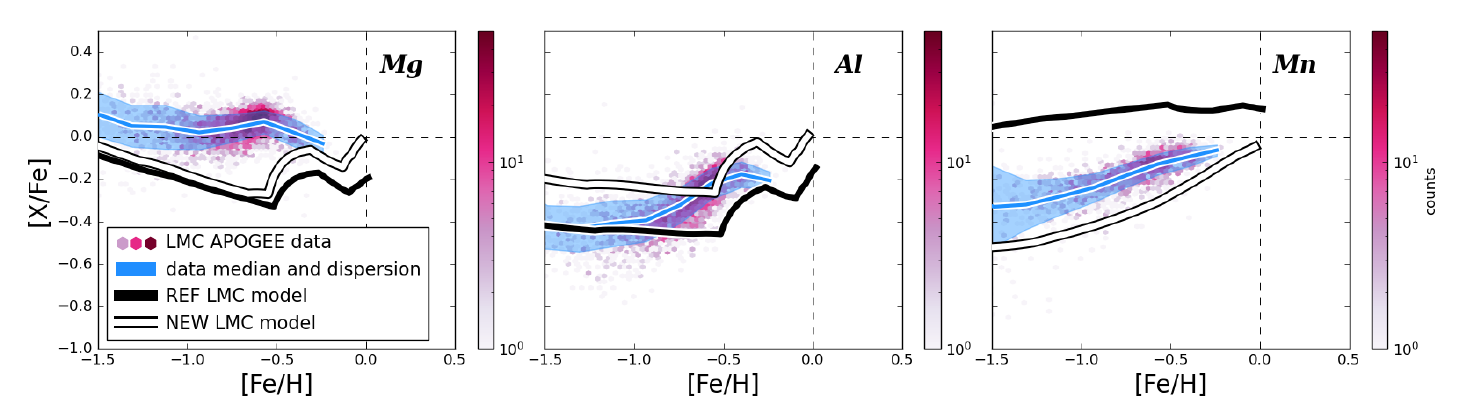}
\caption{LMC abundance ratios of Mg, Al and Mn. The data are from APOGEE DR17, are binned and color coded as in Fig. \ref{fig:MW_abb}. Again, the blue line represents the running median computed with the same bin size and overlap as in Fig. \ref{fig:MW_abb} and the blue area is the standard deviation. The black and the white lines are the REF LMC model and the NEW LMC model, respectively.}
\label{fig:abb_LMC}
\end{figure*}

\subsection{MW: [Mg/Fe], [Al/Fe] and [Mn/Fe]}
\label{subsec:MW_abb}

In Fig. \ref{fig:MW_abb}, we compare  the abundances of Mg, Al and Mn for the Milky Way, as predicted by the REF MW model (black lines) and the NEW MW model (white lines) compared to the APOGEE DR17 data. In the left panel, we note that the REF MW model underestimates Mg throughout the whole [Fe/H] range, which is a quite common feature among many sets of yields for this element \citep[e.g.][]{francois2004,Romano+10,Prantzos+18, lian2020}. The underestimation is particularly pronounced for [Fe/H] $\gtrsim -0.5$ dex and it becomes larger towards the solar metallicity. To fix this behaviour we can add corrections to slightly increase the Mg production, especially at the latest stages of the MW evolution. 
In the middle panel, REF MW model shows an underestimation also for Al. The data show a trend which is quite high at low metallicities ([Al/Fe] $\sim$ 0.25 dex) and then decreases as long as we go towards the solar value. Therefore, Al yield should be increased at low [Fe/H] and then decreased.
In the right panel, we present the Mn abundance. The REF MW model is quite compatible with the range of [Mn/Fe] across the [Fe/H] range spanned by the observations, although the shape of the trend is not that of the running median. In this case the NEW MW model adopts a yield correction that was introduced by \citet{Cescutti+08} with a metallicity dependence, as shown in Table \ref{tab:yields_correction}. This correction actually does not work as well as the original prescriptions but it is worth the investigation also on the [Mg/Mn] vs. [Al/Fe] plane.

Moreover, since the abundances of Mg, Mn and Al seem to be underestimated by the models through the whole [Fe/H] range, together with the corrections to each element already explained we chose to slightly decrease the yield of Fe from massive stars.

The corrections we applied to the adopted yields are shown in Table \ref{tab:yields_correction}. In the first column, we indicate the element analyzed, in the second one we present the correction adopted for the low-intermediate mass stars, in the third one those for massive stars and in the fourth column we list the corrections for the SNIa yield. The check marks represent those cases where no corrections were added. 

For the MIX MW model we assume the NEW MW model prescriptions for Mg, Al and Fe and the REF MW model ones for Mn. We therefore highlight that we do not plot the [Mg/Fe], [Al/Fe] and [Mn/Fe] ratios predicted by the MIX MW model since they are just a combination of the other two models.

Since the MW data set is the largest  available, we consider it to be the benchmark of our work. The corrections presented in Table \ref{tab:yields_correction} will also be tested with a LMC model to check whether they still represent an improvement also for other systems.

\begin{figure*}
    \centering
    \includegraphics[scale=0.2]{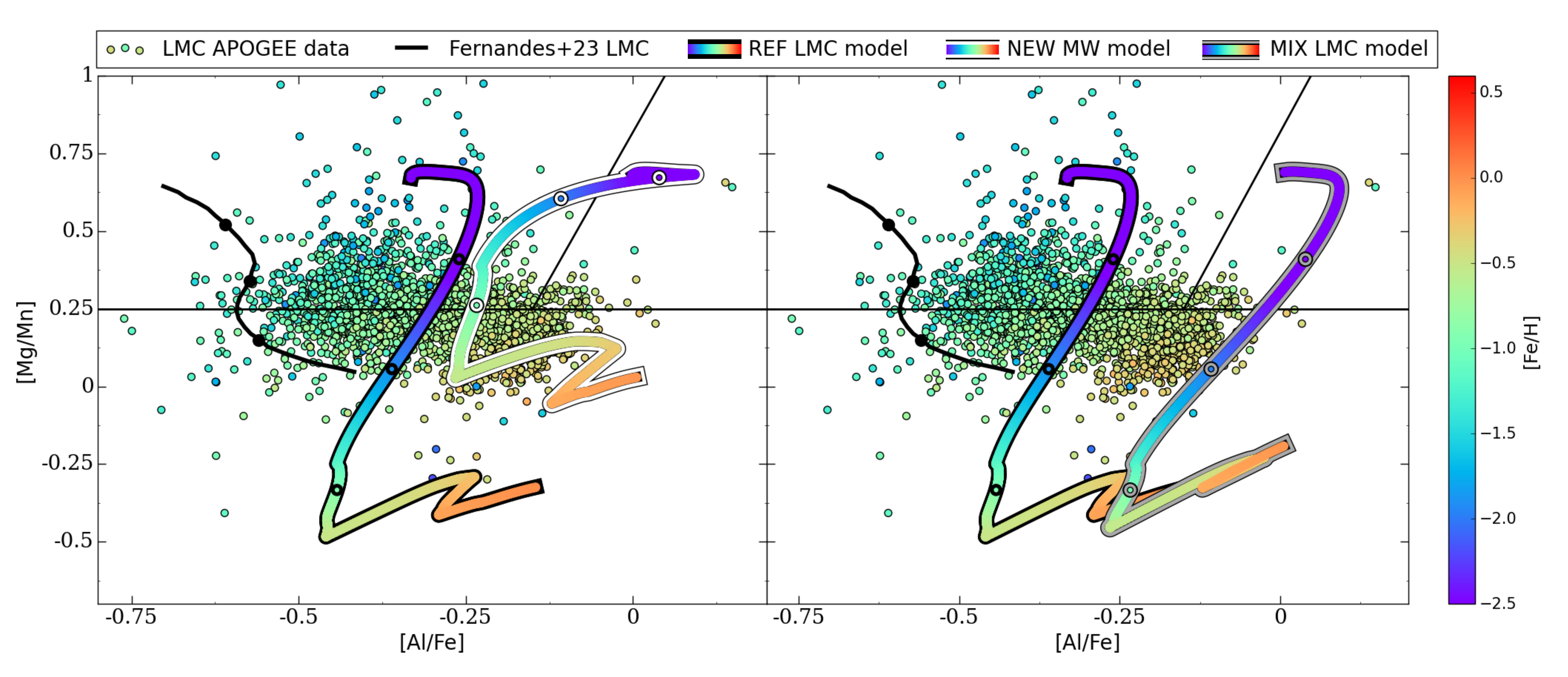}
    \caption{LMC in the [Mg/Mn] vs. [Fe/H] plane. In the two panels we plotted our models compared to the APOGEE DR17 data and to the model from \citet{fernandes2023}. As in Fig. \ref{fig:2_MW_MgMn} the color coding represents the [Fe/H] content. Again, the circles on the models mark three ages, 0.3 Gyr, 1 Gyr and 5 Gyr. Here, unlike Fig. \ref{fig:2_MW_MgMn} the data are not binned. \textit{Left panel}: REF LMC model (color coded black-edge line) and NEW LMC (with yields corrections, color coded white-edge line) are compared to the data and to \citet{fernandes2023} results. \textit{Right panel}: REF LMC model (color coded black-edge line) and MIX LMC model (color coded grey-edge line) together with the data and the \citet{fernandes2023} results.}
    \label{fig:2_LMC_MgMn}
\end{figure*}

\subsection{MW: [Mg/Mn] vs. [Al/Fe]}
\label{subsec:MW_MgMn}
In the two panels of Fig. \ref{fig:2_MW_MgMn}, we show the behaviour of our models in the plane [Mg/Mn] vs. [Al/Fe] compared to the APOGEE DR17 data and to the results by \citet{fernandes2023} obtained with the prescriptions by \citet{andrews2017}. Their model is a one-zone model that assumes a \citet{Kroupa01} IMF, a single infall Schmidt-Kennicutt SF law according to \citet{Kennicutt1998} (as shown in Fig. 6 lower panel of \citealt{fernandes2023}) and yields from \citet{CL04} for CCSNe, \citet{Iwamoto+99} model W70 for SNIa, and \citet{Karakas10} for AGBs. All our models and the data are color-coded according to the [Fe/H] range to show the evolution in metallicity. The data are binned so that the color of each data point represents actually the median [Fe/H] of the data lying in that bin. 
The three circles mark the abundance ratios of the chemical evolution model at 0.3 Gyr, 1 Gyr and 5 Gyr evolutionary times, respectively.

The horizontal and diagonal black lines identify three regions of this plane, each of which hosts a different population of stars. The upper left sector is the locus of the unevolved stars, the upper right hosts the high-$\alpha$ in situ population and the lower region is the low-$\alpha$ in situ population locus. 

In both the panels we show the data, the REF MW model (color coded black-edge line) and the results by \citet{fernandes2023} (black thin line). As the REF MW model did not reproduce the trends of Mg and Al, it does not even reproduce the trend shown by the data in this plane and, in particular, it shows a large underproduction of Al at low [Fe/H], which was also evident in the [Al/Fe] vs. [Fe/H] plot.
On the contrary, the obtained [Fe/H] range is in good agreement with the data, meaning that our choice of applying a small-effect correction on the Fe yield is quite reasonable.

In the left panel we also show the NEW MW model with the yields corrected as in Table \ref{tab:yields_correction} (color coded white-edge model). It is evident that this second model reproduces better the data, matching at the same time the [Fe/H] range. The corrections that have the larger effect on this plane are those on Al and Fe, because their global effect is increasing the [Al/Fe] abundance. The corrections on Mg and Mn yields, on the other side, are less evident in this plane because we increase the production of both Mg and Mn, hence the effect is smoothed out when we study the ratio between them.

In the right panel we show, together with the REF MW model, the MIX MW model. As presented in Table \ref{tab:yields_correction} this model is the combination of the prescription of the REF MW and the NEW MW models. It adopts the corrections for Mg and Al but the original prescriptions for Mn so that we are able to reproduce all the abundances. As it is evident from the right panel we can also fit the [Mg/Mn] vs. [Al/Fe] trend, therefore we can consider this combination of yields as the best one to reproduce the MW abundance features.

\subsection{LMC: [Mg/Fe], [Al/Fe] and [Mn/Fe]}
\label{sec:abb_LMC}
In this section we show the abundances of Mg, Al and Mn in the LMC. For this dwarf galaxy we tested the REF LMC model (black line) and the NEW LMC model (white line), which contains the same yield corrections applied to the MW. Again, we do not show the MIX LMC model since it is just a combination of the other two. We show the abundance ratio predicted compared to the observed ones in Fig. \ref{fig:abb_LMC}.
In this case we see that although none of the two models is able to properly reproduce the abundance patterns observed, the NEW LMC models shows a slight improvement. 
The case of Mn is noteworthy: the corrective factor that in the MW increases the Mn predicted by the model, when applied to the LMC decreases the Mn production substantially. This is easily understood by looking at the Mn corrective factor itself listed in Table \ref{tab:yields_correction}. Since the factor is metallicity dependent its effect on the abundance depends on the environment.

\subsection{LMC: [Mg/Mn] vs. [Al/Fe]}
\label{sec:LMC_MgMn}

In this Section, we show Fig. \ref{fig:2_LMC_MgMn} with the results for the LMC in the plane [Mg/Mn] vs. [Al/Fe] compared to the APOGEE DR17 data for LMC and the previous results presented by \citet{fernandes2023}. The models and the data follow the same color coding as in Fig. \ref{fig:2_MW_MgMn} as well as the division of the plane into three regions and the circles. In this case, the data are not binned since the data set is smaller than for the MW.

The interesting aspect of this plot is the location of the data and the model results on the plane. As already explained, the upper left part of the diagram is the locus of the unevolved populations, where around half of the LMC data points are located. The other half occupies the evolved region closer to the centre of the plot. In this context, the comparison between the LMC and the other dwarf systems nearby is noteworthy. As shown in \citet{fernandes2023} the smallest satellites (e.g. Sculptor, Carina, Fornax, Draco, Sextans and Ursa Minor), which are those with a SFR that was quenched at early times, tend to occupy the leftmost region of the plane whereas the LMC is located more towards the centre of the diagram (with coordinates [Mg/Mn] $\sim$ 0.25 dex and [Al/Fe] $\sim$ -0.10 dex ). This difference could be a consequence of the SFH of the LMC, which presents active star formation at the present time, at variance with the other dwarf spheroidal galaxies. In our models, we adopted this kind of SFR and we were able to reproduce the observed evolution towards the centre of the plot. On the contrary, the Milky Way has a present day star formation lower than the initial peak, and this produces an opposite evolution, from the right to the left of the plot.

As in Fig. \ref{fig:2_MW_MgMn} we show two panels displaying the models adopted. On both panels we show the data compared to the REF LMC model (color coded black-edge line) and to the results by \citet{fernandes2023} (black thin line) based on the LMC prescriptions by \citet{andrews2017}. It is evident that this model cannot reproduce the observations due to a too small Mg abundance and a too large Mn abundance. On the left panel we compare these two trends with the NEW LMC model. This model does not reproduce the data either, but it represents an improvement with respect to the reference one. In particular, as shown in Fig. \ref{fig:2_LMC_MgMn}, the [Mg/Mn] predicted is substantially improved.

On the right panel, we compare the REF LMC model to the MIX LMC model, that for the MW resulted to be the best one.

To highlight from a quantitative point of view the best model among the three tested we computed the average [Mg/Mn] weighted on the stars formed in each timestep and compared it to the average [Mg/Mn] obtained from the data. REF MW model has an average [Mg/Mn] of $\sim$ -- 0.32 dex, for the NEW LMC model the average [Mg/Mn] is $\sim$ 0.16 dex, and for the MIX LMC model it is $\sim$ -- 0.29. The data, on the other side, show an average [Mg/Mn] of $\sim$ 0.23, which is in much better agreement with the result from the NEW LMC model.

The reason of the behaviour of the different models lies in the Mn yield. In the case of the MW the original Mn prescription was the best one to reproduce the data whereas for the LMC, even though none of the models overlaps the running median, the new prescriptions work slightly better. Therefore the NEW LMC model is the best one to describe the LMC.

\subsection{Testing different nucleosynthesis and IMF prescriptions}
\label{sec:MW_andrews}

\begin{figure*}
\centering
\includegraphics[scale=1.5]{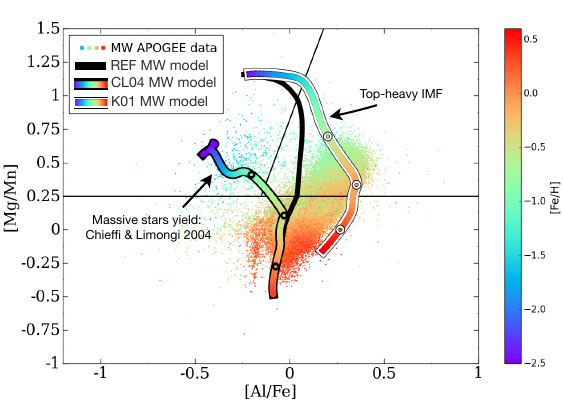}
\caption{[Mg/Mn] versus [Al/Fe] for the CL04 MW model (color coded line, black edge) and the K01 MW model (color coded line, white edge) compared to the APOGEE DR17 data and the REF MW model (black line). The color coding of the data, the CL04 MW model and the K01 MW model is the same as Fig. \ref{fig:2_MW_MgMn} as well as the binning of the data and the three circles plotted on the model. }
\label{fig:CL04_K01}
\end{figure*}

In \citet{fernandes2023}, they compared APOGEE DR17 data with the predictions obtained by the chemical evolution models of \citet{andrews2017}. Therefore, in addition to our models, we also test their nucleosynthesis prescriptions: they use the same sets of yields except for massive stars, where they adopt \citet{CL04}. We test this combination of yields for the Milky Way in order to show how sensitive the [Mg/Mn] versus [Al/Fe] plane is to the nucleosynthesis prescriptions. We label this test model as CL04 MW model and we show the results compared to the data and to the REF MW model in Fig. \ref{fig:CL04_K01}. Here we can see that, only by changing the yields for the massive stars and no other parameters, we obtain a behaviour which is substantially different from the reference case shown in Fig. \ref{fig:2_MW_MgMn}. Since we modified the production from massive stars, the differences concern the first part of the evolution of the Galaxy due to the massive star explosion timescales. Nevertheless, it is evident how we cannot reproduce the abundance pattern nor the [Fe/H] range.

The reason for this is that the nucleosynthesis of the elements considered in this plot is quite uncertain. Mg is known to be underestimated by many different sets of yields, the nucleosynthesis of Mn is quite uncertain (being Mn a Fe-peak element), and that of Al has a dependence on the metallicity that introduces a further level of complexity. Hence, the adoption of the ratios of the above elements makes it quite difficult to draw firm conclusions. By changing the yields of these elements within their uncertainties one can obtain trends on the [Mg/Mn] vs. [Al/Fe] plot with different behaviours, slopes and [Fe/H] ranges. 

In principle, the combination of these elements would allow us to disentangle some properties of the galaxies analyzed, such as the SFR, but given the state of the art of the yields available in the literature, the nuclear uncertainties prevent this plane to be a strong diagnostic for the star formation history of a galaxy.

We performed a second comparison with \citet{fernandes2023} adopting all the parameters of REF MW model except the IMF, for which we adopted the one they used, namely \citet{Kroupa01}. We show this model in Fig. \ref{fig:CL04_K01}. The \citet{Kroupa01} IMF is a top-heavy IMF which produces more massive stars than the \citet{Kroupa1993} that we adopted in the REF MW model and that was derived for the MW. The  \citet{Kroupa01}  IMF  produces a more rapid increase of the metallicity than the \citet{Kroupa1993} one, and the net result is  an increase of the [Al/Fe] ratios and the range in [Fe/H], relative to the other models, as highlighted by the color coding in the figure. 

The reason for choosing  the \citet{Kroupa1993} IMF instead of the more recent \citet{Kroupa01} one, is that the former was specifically derived for the solar vicinity,  while the latter is a sort of universal IMF. Many previous papers on the MW  have suggested the \citet{Kroupa1993} IMF as the best one to reproduce the main features of the Galactic disc (see for example \citealt{Romano+10}). For external galaxies, such as the satellites of the MW, the situation is different because no observational derivation of the IMF exists. In these cases, normally a Salpeter IMF is used, and this IMF is similar to the \citet{Kroupa01} one.

\subsection{The effects of Galactic wind}

\begin{figure}
\centering
\includegraphics[scale=0.17]{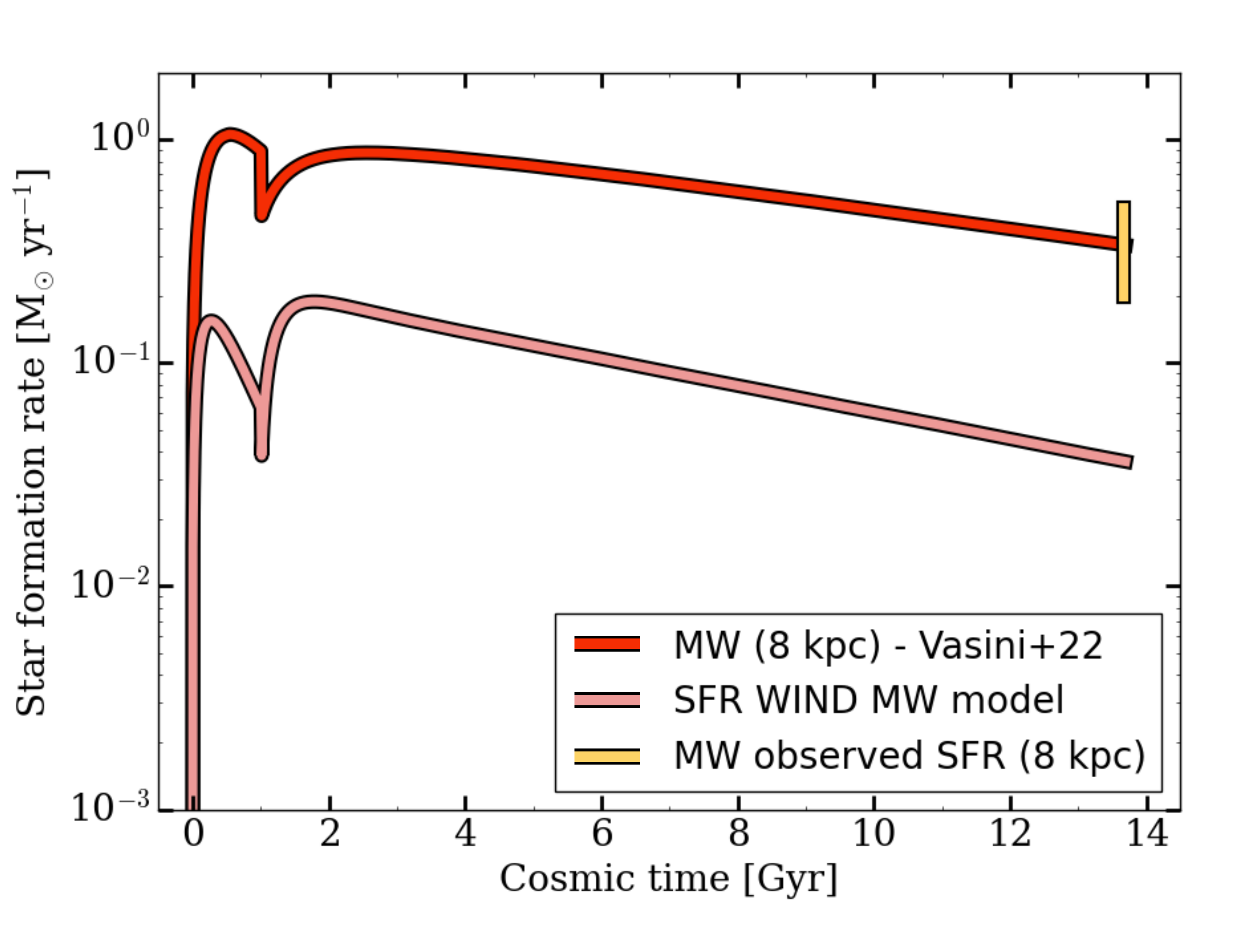}
\caption{Star formation histories for the MW models without the wind contribution (red line) and for the WIND MW model (pink line) at 8 kpc from the Galctic centre, compared to the observed data (see reference in Fig. \ref{fig:MW_LMC_SFR}).}
\label{fig:SFR_wind}
\end{figure}

\begin{figure*}
\centering
\includegraphics[scale=1.5]{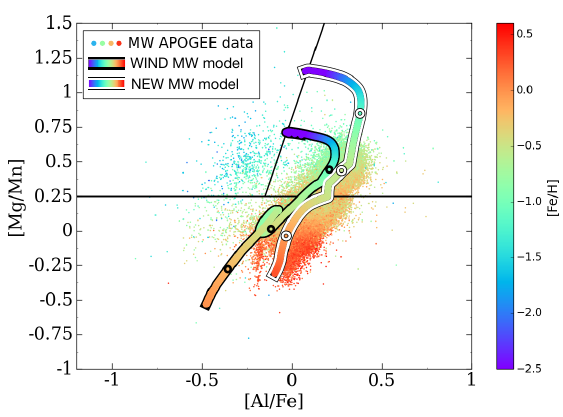}
\caption{[Mg/Mn] versus [Al/Fe] for the WIND MW model (color coded line, black edge) and the NEW MW model (color coded line, white edge) compared to the APOGEE DR17 data. The color coding is the same as Fig. \ref{fig:2_MW_MgMn} as well as the binning of the data and the circles plotted on the model. }
\label{fig:Wind_ref}
\end{figure*}

 In  \citet{fernandes2023}, they compared APOGEE data with the predictions of the chemical evolution model proposed by \citet{andrews2017},  where a  Galactic wind   is assumed for the MW.  In our REF MW and NEW MW models we do not include any outflow since, for a potential well such that of the MW, only Galactic fountains are allowed \citep{melioli2008,melioli2009,spitoni2008,spitoni2009}. Therefore, for comparison with \citet{andrews2017}
 model, we computed  a model for the MW with Galactic wind (WIND MW model, see Table \ref{tab:model_parameters}). 
 The introduction of the wind term into the chemical evolution equation works as a depletion term that decreases the gas available for the star formation. The net effect is that of a decline in the star formation rate (see Fig. \ref{fig:SFR_wind}), with less stars formed and therefore a lower abundance of metals restored into the interstellar medium. The wind affects the abundances in a very clear way, since the abundance ratios of elements produced on different time-scales, e.g. [Mg/Mn], (in the case of a wind independent of the chemical species) drop drastically,  because the effect of the gas loss is immediate on the elements produced on short timescales (e.g. Mg) and therefore related to the SFR. Mathematically, the intensity of the wind depends on the mass loading factor $\omega_{i}$, as shown in Sec. \ref{sec:model_LMC}, that in our case we imposed equal to 2.5, as done by \citet{andrews2017}. Regarding the other prescriptions the WIND MW model follows the NEW MW model.
 In Fig. \ref{fig:Wind_ref} we show that the WIND MW model that in the [Mg/Mn] vs. [Al/Fe] plane is shifted towards smaller values of both the abundance ratios. In addition, the range of [Fe/H] spanned by the WIND MW model is consistently lower than that spanned by the data, as shown by the color coding. In conclusion, once we add the wind to our models we are not anymore able to properly reproduce the data trend or the range in [Fe/H].

\section{Conclusions}
\label{conclu_sec}
In this work, we studied the chemical evolution of the Milky Way and of the Large Magellanic Cloud in order to understand the role of the [Mg/Mn] vs. [Al/Fe] plane in unveiling the star formation history of  galaxies. According to the observational data, different stellar populations belonging to the same galaxy lie in different regions of the plane (as explained in Sec. \ref{subsec:MW_MgMn}). Hence, the galaxy pattern on this diagram is the result of the imprints left on those abundances by the SFH of that specific galaxy. In principle, this element combination is ideal to extract information about the SFH, since they have different sites and time scales of production, but a discussion on the related uncertainties was  still missing in the literature. Our aim was to analyze the uncertainties, in particular those related to the nucleosynthesis of these elements. 

We did that by the means of three already tested models for the Milky Way and three for the LMC, where we varied the nucleosynthesis prescriptions. We also introduced a fourth model for the Milky Way adopting the same yields from massive stars as in Andrews et al. (2017), with the aim of showing how the trends on the [Mg/Mn] vs. [Al/Fe] plane depend on the assumed stellar nucleosynthesis.


Analysing the [Mg/Mn] vs. [Al/Fe] for the different models presented, our results can be summarised as follows: 

\begin{itemize}
    \item to reproduce the abundance patterns of Mg, Al and Mn in the MW as shown by APOGEE DR17 data, it is necessary to add corrective factors to the yields adopted in \citet{Vasini+22}. In particular, as shown in Table \ref{tab:yields_correction}, we modified the massive stars contribution for Fe, Mg and Al, the low-intermediate mass star contribution for Al and the SNIa production for Mn.

    \item Regarding the LMC [Mg/Mn] vs. [Al/Fe] diagram, the corrected yields improve the agreement with the APOGEE data also for this galaxy. This means that, even if the corrective factors where tailored on the Milky Way they work also for the LMC. At the same time the recent SFR bursts, assumed for the LMC, account for those stars found on the evolved star region.

    \item Changing the yields for massive stars only, as shown in Fig. \ref{fig:CL04_K01}, produces completely different results. By assuming different sets of yields one can obtain very different behaviours of the [Mg/Mn] vs. [Al/Fe], as well as the predicted range of [Fe/H]. 
    In the same way, also changing the IMF by adopting a top-heavy one modifies the behaviour on the [Mg/Mn] vs. [Al/Fe] plane predicting larger [Al/Fe] ratios, since the fraction of massive stars, which produce Al,  is larger.
    In the paper of \citet{andrews2017}, who computed the same diagram for the MW and satellite galaxies,  the yields of CL04, the \citet{Kroupa01} IMF and the presence of galactic  winds in the MW are adopted. Their predicted trends are marginally fitting the data and in some cases are different from ours. In our standard model, we did not adopt the galactic winds in the MW, since it is more probable that in our Galaxy the gas produces Galactic fountains rather than winds, due to the deep potential well. However, we also computed a model including the Galactic wind with the same parameters as in \citet{andrews2017} and the results produce a worse agreement with data than our standard model without wind.
      
    \item Unfortunately,  no firm conclusions are possible on the basis of theoretical models, mainly because of the uncertainties still present in the stellar yields and the IMF in our Galaxy and in its satellites, the two quantities that influence most the abundance ratios. 
     \end{itemize}
     
    In conclusion, this work focuses on  an issue which is not related only to the [Mg/Mn] vs. [Al/Fe] plane but also to all the other chemical elements. Many sets of stellar yields are available in the literature but they differ from one another, due to different nucleosynthesis and stellar evolution prescriptions and the presence or absence of mass loss and rotation.  In addition, there are some elements whose nucleosynthesis is very poorly known, and among those elements there are indeed Mn and Al. 
    Therefore, it is risky to rely only on one diagram such as that discussed in this paper. Other chemical clocks should be adopted at the same time, such as for example s-process elements  and alpha-elements relative to Fe, coupled with the kinematic properties.

\section*{Acknowledgements}
The authors kindly thank the referee for all the interesting comments and suggestions that improved substantially the paper.  We also thank I.N.A.F. for the 1.05.12.06.05 Theory Grant
- Galactic archaeology with radioactive and stable nuclei.

\bibliographystyle{aa} 
\bibliography{arianna_mgmn}

\end{document}